\pdfoutput=1

\documentclass[11pt]{article}

\usepackage[]{acl}

\usepackage{times}
\usepackage{latexsym}

\usepackage[T1]{fontenc}

\usepackage[utf8]{inputenc}

\usepackage{microtype}

\usepackage{multirow}
\usepackage{arydshln} 
\usepackage{adjustbox}

\usepackage{xcolor} 
\usepackage{colortbl}

\usepackage{todonotes}

\newcommand{\conv}{\textsc{\small \textbf{ConvFeat} }}
\newcommand{\gemaps}{\textsc{\small \textbf{eGeMAPSv02} }}
\newcommand{\wav}{\textsc{\small \textbf{wav2vec} }}

\title{Impact of Environmental Noise on Alzheimer's Disease Detection\\ from Speech: Should You Let a Baby Cry?}

\author{Jekaterina Novikova \\
  Winterlight Labs / Toronto, Canada \\
  \texttt{jekaterina@winterlightlabs.com} \\}

\begin{document}
\maketitle
\begin{abstract}
Research related to automatically detecting Alzheimer's
disease (AD) is important, given the high prevalence of AD and
the high cost of traditional methods. Since AD significantly
affects the acoustics of spontaneous speech, speech processing and machine learning (ML) provide promising techniques for reliably detecting AD. However, speech audio may be affected by different types of background noise and it is important to understand how the noise influences the accuracy of ML models detecting AD from speech.
In this paper, we study the effect of fifteen types of environmental noise from five different categories on the performance of four ML models trained with three types of acoustic representations. We perform a thorough analysis showing how ML models and acoustic features are affected by different types of acoustic noise. We show that acoustic noise is not necessarily harmful - certain types of noise are beneficial for AD detection models and help increasing accuracy by up to 4.8\%. We provide recommendations on how to utilize acoustic noise in order to achieve the best performance results with the ML models deployed in real world. 
\end{abstract}

\section{Introduction}

 Alzheimer's disease (AD) is a progressive neurodegenerative disease that affects over 40 million people worldwide~\cite{prince2016world}. Current forms of diagnosis are both time consuming and expensive \cite{prabhakaran2018analysis}, which might explain why almost half of those living with AD do not receive a timely diagnosis~\cite{jammeh2018machine}. Studies have shown that ML methods can be applied to distinguish between speech from healthy and AD participants~\cite{fraser2016linguistic,balagopalan2018effect,zhu2019detecting,eyre2020fantastic}. Currently, speech recording for AD-related research typically takes place in
a quiet room with a guiding clinician. Given that smartphone technology is rapidly
advancing, speech assessments using ML models trained on recordings obtained by smartphones offer a potentially simple-to-administer and inexpensive solution, scalable to the
entire population, that can be performed anywhere, including the patient's home~\cite{kourtis2019digital,mc2019can,fristed2021evaluation}. However, the problem of model robustness to acoustic noise becomes increasingly important when deploying ML models in real world~\cite{robin2020evaluation}.

Current popular approaches to dealing with acoustic noise in AD detection models involve: 1) eliminating noise using various audio pre-processing techniques~\cite{luz2021detecting}, 2) selecting features that are resilient to ASR error/noise~\cite{zhou2016speech}, 3) minimizing the effects of noise with multimodal fusion of features~\cite{rohanian2021alzheimer}. All these approaches share a common assumption of acoustic noise being definitely harmful for ML models detecting AD from speech. However, in other ML research areas, such as computer vision or NLP, adding a certain level of natural and artificial noise to data is considered a valid and advantageous practice that helps achieving better performance in tasks like image recognition~\cite{koziarski2017image,steffens2019can}, text generation~\cite{feng2020genaug} and relation classification~\cite{giridhara2019study}, among others. The recent studies in AD classification from transcribed speech show that small levels of linguistic noise do not negatively affect performance of BERT-based models~\cite{novikova2021robustness}, although there is a difference in predictive power between lexical and syntactic features, when it comes to AD detection from speech~\cite{novikova2019lexical}. 

Motivated by the previous work, in this paper we study the effect of acoustic noise on performance of the ML models trained to detect AD from speech. The contributions of this paper are: 

\begin{enumerate}
    \item we analyze the effect of environmental acoustic noise on the values of acoustic features extracted from speech; 
    \item we perform a thorough study on the effect of acoustic noise on AD classification performance across ML models, extracted acoustic features and noise categories; 
    \item we provide recommendation to ML researchers and practitioners on how to utilize acoustic noise in order to achieve the best performance results.
\end{enumerate}

\section{Related Work}

\subsection{Environmental Noise and Speech Quality}


Multiple previous studies attempted to investigate the influence of the environment background noise on speech quality. For example, \citet{naderi2018speech} conducted a study in which participants rated the quality of speech files first in the laboratory and then in noisy speech collection settings, such as cafeteria and living room. They found that the presence of a ``cafeteria" or a ``crossroad" background noise would decrease the correlation to speech quality ratings. 

Furthermore, multiple studies have addressed the issue of speech intelligibility under certain background noise conditions. To name some, \citet{meyer2013speech} tackled the problem of speech recognition accuracy in ecologically valid natural background noise
scenarios and showed the relation between the levels of noise and confusion of vowels, lexical identification and perceptual consonant confusion. 

\citet{jimenez2020effect}  investigated the influence of environmental background noise on speech quality, where the quality of speech files was assessed under the influence of two types of background noise at different levels, i.e., street noises and tv-show. The authors found there was a certain threshold of the environment background noise level that impacted the quality of speech, and different types of noise had a different effect on the quality.

Motivated by the previous studies, in this work we analyze fifteen different types of environmental background noise in order to figure out differences in their impact. We also compare the impact of short and continuous noise to follow up on the findings of the impact threshold. 

\subsection{Alzheimer's Disease Detection in Noisy Settings}

Given the number of people with AD is growing and the population is aging fast in many countries \cite{brookmeyer2018forecasting}, it becomes more and more important to have tools to help identify the presence of cognitive impairment relating to AD that can be deployed frequently, and at scale. This need will only increase as effective interventions are developed, requiring the ability to identify patients early in order to facilitate prevention or treatment of disease \cite{vellas2021new}. 
Most of the current AD screening tools represent a significant burden, requiring invasive procedures, or intensive and costly clinical testing. 
However, recent shifts toward telemedicine and increased digital literacy of the aging population provide an opportunity for using digital health tools that are ideally poised to meet the needs for novel solutions. 
Recently, automated tools have been developed that assess speech and can be used on a smartphone or tablet, from one's home \cite{robin2021using}.
Digital assessments that can be accessed on a smartphone or tablet, completed from home and periodically repeated, would vastly improve the accessibility of AD screening compared to current clinical standards that require clinical visits, extensive neuropsychological testing or invasive procedures. 

The pervasiveness of high-quality microphones in smart devices makes the recording of speech samples straightforward, not requiring additional equipment or sensors. However, there is a lack of control over the participants performing digital assessments in home environment, and often not enough information is collected about their playback system and background environment. Participants might be exposed to different environmental conditions while executing specific tasks, and as such, their recorded speech quality may be disturbed with some background noise.

In the speech community, the active ongoing effort is focused on solving the problem of automated speech enhancement with the methods  of noise suppression that are based on machine learning and deep learning (\citealp{zhang2022multi,braun2021towards,choi2018phase,odelowo2017noise}, among many others). However, this problem is not considered to be solved, and the research community continues developing methods for effective noise elimination from audio recordings \cite{dubey2022icassp}.

These challenges motivate us asking a question whether noise suppression is absolutely necessary when it comes to the specific task of AD detection from speech. In this work, we perform a thorough study on the effect of acoustic background noise, standard for home environments, on AD classification performance across a range of ML models.

\subsection{Speech Quality and Alzheimer's Disease Detection}

Speech is a promising modality for digital assessments of cognitive abilities. Producing speech samples is a highly ecologically valid task that requires little instruction and at the same time is instrumental to daily functioning. Advances in signal processing and natural language processing have enabled objective analysis of speech samples for their acoustic properties, providing a window into monitoring motor and cognitive abilities. Most importantly, previous research has extensively shown that speech patterns are affected in AD, demonstrating the clinical relevance of speech for detecting cognitive impairment and dementia \cite{martinez2021ten,de2020artificial,slegers2018connected}.

Some of the features employed to describe acoustic characteristics of the voice applied to AD detection, include conventional acoustic features, such as fundamental frequency, jitter and shimmer, as well as pre-trained embeddings from deep neural models
for audio representation, such as wav2vec \cite{balagopalan2021bcomparing}. 
Quality of speech, which may be influenced by environmental noise, inevitably affects the values of these acoustic features extracted from speech and as a result may potentially influence the performance of ML models that use these features as internal representations of human speech.

However, in other research areas, such as computer vision or NLP, adding a certain level of natural or artificial noise to data is considered a valid and advantageous practice that helps achieving better performance in tasks like image recognition~\cite{koziarski2017image,steffens2019can}, text generation~\cite{feng2020genaug} and relation classification~\cite{giridhara2019study}, among others. Moreover, deep neural acoustic models, such as wav2vec, that are used to generate acoustic embeddings used in AD detection, are pre-trained on healthy speech. As such,  it is possible that the subparts of the embeddings that are affected by environmental noise are not used for the task of AD detection directly and as a result, they do not influence the performance of such detection.

In this work, we make an attempt to understand how different types of environmental noise are impacting the values of different types of acoustic features extracted from speech, as well as how this affects performance of ML models relying on these features.

\section{Methodology}

\subsection{Dataset} 

\begin{figure}[t!]
    \centering
    \includegraphics[width = \linewidth]{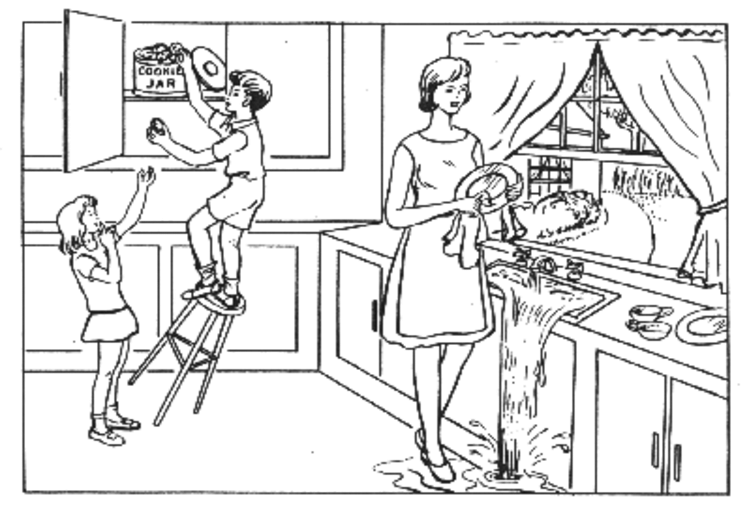}
    \caption{Cookie Theft picture used to collect speech for the ADReSS\textit{o} dataset.}
    \label{fig:cookie-theft}
\end{figure}

We use the ADReSS\textit{o} Challenge dataset \cite{luz2021detecting}, which consists of 166 training speech samples from non-AD (N=83) and
AD (N=83) English-speaking participants. Speech is elicited from participants through the Cookie Theft picture from the Boston Diagnostic Aphasia exam (Figure~\ref{fig:cookie-theft}). In contrast to the
other datasets for AD detection such as DementiaBank’s English Pitt Corpus,
the ADReSS\textit{o} challenge dataset is well balanced in terms of age and gender. In addition,
the pre-processing step of ADReSS\textit{o} recordings were acoustically enhanced with stationary noise removal and audio volume normalisation applied across all speech segments to control for variation caused by recording conditions such as microphone placement. Such enhancements make this dataset a great source of the noise-clean audio, which is important for our experiments.

\subsection{Feature Extraction} 
The following groups of features were extracted for the further use in the experiments:

\begin{enumerate}
    \item \conv : We extract 182 acoustic features from the unsegmented speech audio files. Those include several statistics such as mean, std, median, etc. of mel-frequency cepstral coefficients (MFCCs), onset detection, rhythm, spectral and power features, following prior works in AD classification~\cite{fraser2016linguistic,zhu2018semi,balagopalan2020bert}.

\item \gemaps :  The extended Geneva Minimalistic Acoustic Parameter Set (eGeMAPS) features are a selected standardized set of statistical features that characterize affective
physiological changes in voice production. We extracted these features for the entire recording, as this feature set was shown to be usable for atypical speech~\cite{xueacoustic} and was successfully used for classifying AD from speech~\cite{gauder21_interspeech,pappagari21_interspeech}.

\item \wav : In order to create audio representations using this approach, we make use of the huggingface\footnote{https://huggingface.co/models} implementation of the wav2vec~2.0~\cite{baevski2020wav2vec} base model \textit{wav2vec2-base-960h}. This base model is pretrained and fine-tuned on 960 hours of Librispeech on 16kHz sampled speech audio. Closely following~\cite{balagopalan2021bcomparing} that used these representations for AD classification, we extracted the last hidden state of the wav2vec2 model and used it as an embedded representation of audio.  
\end{enumerate}

\begin{table*}[t!]
\begin{adjustbox}{max width=\linewidth, center}
\begin{tabular}{lllcc}
 &  &  & \multicolumn{2}{l}{\textbf{Ratio of sign diff features}} \\
\begin{tabular}[c]{@{}l@{}}\textbf{Noise}\\\textbf{category}\end{tabular} & \textbf{Subcategory} & \textbf{Features} & \begin{tabular}[c]{@{}l@{}}\textbf{Short}\\ \textbf{noise}\end{tabular} & \begin{tabular}[c]{@{}l@{}}\textbf{Background}\\ \textbf{noise}\end{tabular} \\
\hline \hline
\multirow{6}{*}{Animals} & \multirow{2}{*}{cat} & \conv & 32.42\% & \cellcolor{red!18}68.68\% \\
 &  & \gemaps & 32.95\% & \cellcolor{red!18}50.00\% \\
 \cdashline{2-5}
 & \multirow{2}{*}{crow} & \conv & \cellcolor{red!18}55.49\% & \cellcolor{red!50}80.22\% \\
 &  & \gemaps & 45.45\% & \cellcolor{red!18}59.09\% \\
\cdashline{2-5}
& \multirow{2}{*}{dog} & \conv & 23.08\% & \cellcolor{red!18}63.19\% \\
 &  & \gemaps & 23.86\% & \cellcolor{red!18}50.00\% \\
 \hline
\multirow{6}{*}{Natural} & \multirow{2}{*}{\begin{tabular}[c]{@{}l@{}}chirping\\birds\end{tabular}} & \conv & \cellcolor{red!18}69.23\% & \cellcolor{red!18}71.43\% \\
 &  & \gemaps & 44.32\% & \cellcolor{red!18}54.55\% \\
 \cdashline{2-5}
 & \multirow{2}{*}{rain} & \conv & \cellcolor{red!18}67.58\% & \cellcolor{red!50}90.11\% \\
 &  & \gemaps & 32.95\% & \cellcolor{red!18}69.32\% \\
 \cdashline{2-5}
 & \multirow{2}{*}{wind} & \conv & 48.35\% & \cellcolor{red!18}78.02\% \\
 &  & \gemaps & 42.05\% & \cellcolor{red!18}60.23\% \\
 \hline
\multirow{6}{*}{Human} & \multirow{2}{*}{coughing} & \conv & 37.36\% & \cellcolor{red!18}52.20\% \\
 &  & \gemaps & 27.27\% & 32.95\% \\
 \cdashline{2-5}
 & \multirow{2}{*}{\begin{tabular}[c]{@{}l@{}}crying\\baby\end{tabular}} & \conv & \cellcolor{red!18}53.30\% & \cellcolor{red!18}68.68\% \\
 &  & \gemaps & 40.91\% & \cellcolor{red!18}67.05\% \\
 \cdashline{2-5}
 & \multirow{2}{*}{sneezing} & \conv & 10.44\% & 41.21\% \\
 &  & \gemaps & 27.27\% & 25.00\% \\
 \hline
\multirow{6}{*}{\begin{tabular}[c]{@{}l@{}}Domestic/\\interior\end{tabular}} & \multirow{2}{*}{\begin{tabular}[c]{@{}l@{}}clock\\ticking\end{tabular}} & \conv & 48.35\% & \cellcolor{red!18}63.74\% \\
 &  & \gemaps & 23.86\% & 30.68\% \\
 \cdashline{2-5}
 & \multirow{2}{*}{\begin{tabular}[c]{@{}l@{}}vacuum\\cleaner\end{tabular}} & \conv & \cellcolor{red!18}63.19\% & \cellcolor{red!50}87.36\% \\
 &  & \gemaps & 42.05\% & \cellcolor{red!18}60.23\% \\
 \cdashline{2-5}
 & \multirow{2}{*}{\begin{tabular}[c]{@{}l@{}}washing\\machine\end{tabular}} & \conv & \cellcolor{red!18}51.10\% & \cellcolor{red!50}82.97\% \\
 &  & \gemaps & 28.41\% & \cellcolor{red!18}65.91\% \\
 \hline
\multirow{6}{*}{\begin{tabular}[c]{@{}l@{}}Urban/\\exterior\end{tabular}} & \multirow{2}{*}{\begin{tabular}[c]{@{}l@{}}car\\horn\end{tabular}} & \conv & 39.01\% & \cellcolor{red!50}81.32\% \\
 &  & \gemaps & 27.27\% & 45.45\% \\
 \cdashline{2-5}
 & \multirow{2}{*}{siren} & \conv & \cellcolor{red!18}53.30\% & \cellcolor{red!18}74.73\% \\
 &  & \gemaps & 42.05\% & \cellcolor{red!18}62.50\% \\
 \cdashline{2-5}
 & \multirow{2}{*}{train} & \conv & \cellcolor{red!18}56.04\% & \cellcolor{red!50}83.52\% \\
 &  & \gemaps & 39.77\% & \cellcolor{red!18}57.95\%
\end{tabular}
\end{adjustbox}
\caption{Impact of noise addition on the value of \conv and \gemaps. Ratio of sign. diff. features shows the percentage of all the features that is significantly ($p<0.05$) different from the original values as a result of adding short noise and background noise to original audio samples. Lighter cell color indicates higher than 50\% ratio, darker - higher than 80\% ratio.}
\label{tab:feature-value-change}
\end{table*}

\subsection{Adding Noise}

We used the audiomentations\footnote{https://github.com/iver56/audiomentations} library to add two types of audio noise that are common when recording audio with smart devices - 1) background noise, and 2) short noise \cite{vhaduri2019nocturnal,dibbo2021effect}. We use a reduced version of the ESC-50 dataset \cite{piczak2015esc} to generate noisy audio, where we select three classes of noise from all the five presented major categories:

\begin{enumerate}
\item\textbf{Animal} sounds: dog, cat, crow
\item\textbf{Natural} soundscapes: rain, wind, chirping birds
\item\textbf{Human} sounds: crying baby, sneezing, coughing
\item\textbf{Domestic / interior} sounds: clock ticking, washing machine, vacuum cleaner
\item\textbf{Urban / exterior} noises: train, car horn, siren
\end{enumerate}

\subsection{Experiments}

We first analyze how significantly addition of noise changes the values of acoustic features \conv and \gemaps. We calculate the ratio of features that are impacted significantly by noise, with the Mann–Whitney U test used to estimate significance of difference.

Next, we experiment with the effect of noise addition to the performance of AD classification models. Following multiple previous studies on AD classification from speech~ \cite{balagopalan2020bert,balagopalan2021acomparing,balagopalan2021bcomparing}, we use a set of linear and non-linear ML models: Logistic regression (LR), Support Vector Machines (SVM), Neural Network (NN), and Decision Tree (DT). 

We use 10-fold cross-validation approach to evaluate the performance of classifiers, with the F1 score being the main classification performance evaluation metric. 

\section{Results and Discussion}

\subsection{Effect of Noise on the Values of Acoustic Features}
\label{sec:noise-effect-values}

The results in Table~\ref{tab:feature-value-change} show that different types of noise have very different impact on the acoustic features, where \textit{sneezing} sound introduced several times within recordings for short periods only affects 10\% of \conv, while continuous background sound of rain significantly changes more than 90\% of these features. Unsurprisingly, background noise affects recordings much stronger than short noise. Notably, conventional acoustic features are on average more vulnerable than \gemaps to both short noise (12.5\% higher ratio of significantly affected features) and background noise (19.8\% higher ratio), with the categories of \textit{natural sounds, domestic/interior} and \textit{urban/exterior} bringing the strongest difference between the \conv and \gemaps. 

Both \conv and \gemaps are quite robust to the \textit{human} non-speech noise, especially the sound of \textit{sneezing}. Out of all the noise types analyzed in this work, \textit{sneezing} is the only one that only affects up to 50\% of acoustic features, both in a format of short and background noise. \textit{Natural} sounds, such as \textit{rain}, \textit{wind} or \textit{chirping birds}, affect the acoustic features the strongest. 

The above results suggest that noise strongly disturbs the quality of audio samples, as represented by both \conv and \gemaps. Next, we analyze whether such a disturbance is beneficial or harmful when it comes to AD detection from disturbed speech.

\subsection{Effect of Noise on Performance of AD Classification}
\label{sec:noise-clf-perf}
Four types of ML models (SVM, neural network / NN, logistic regression / LR and decision tree / DT) were trained on noisy and original audio recordings represented using \conv, \gemaps and \wav. Each set of features was extracted from both original audio recordings and the recordings with added 20 subcategories of noise. Each ML model was evaluated with the F1 score on three different random seeds. As such, it is possible to analyse the mean classification performance level per feature type, where performance is averaged across all the seeds, for each model, noise subcategory and feature type. 

\subsubsection{Analysis Per Feature Type}
\begin{figure}[t!]
    \centering
    \includegraphics[width = \linewidth]{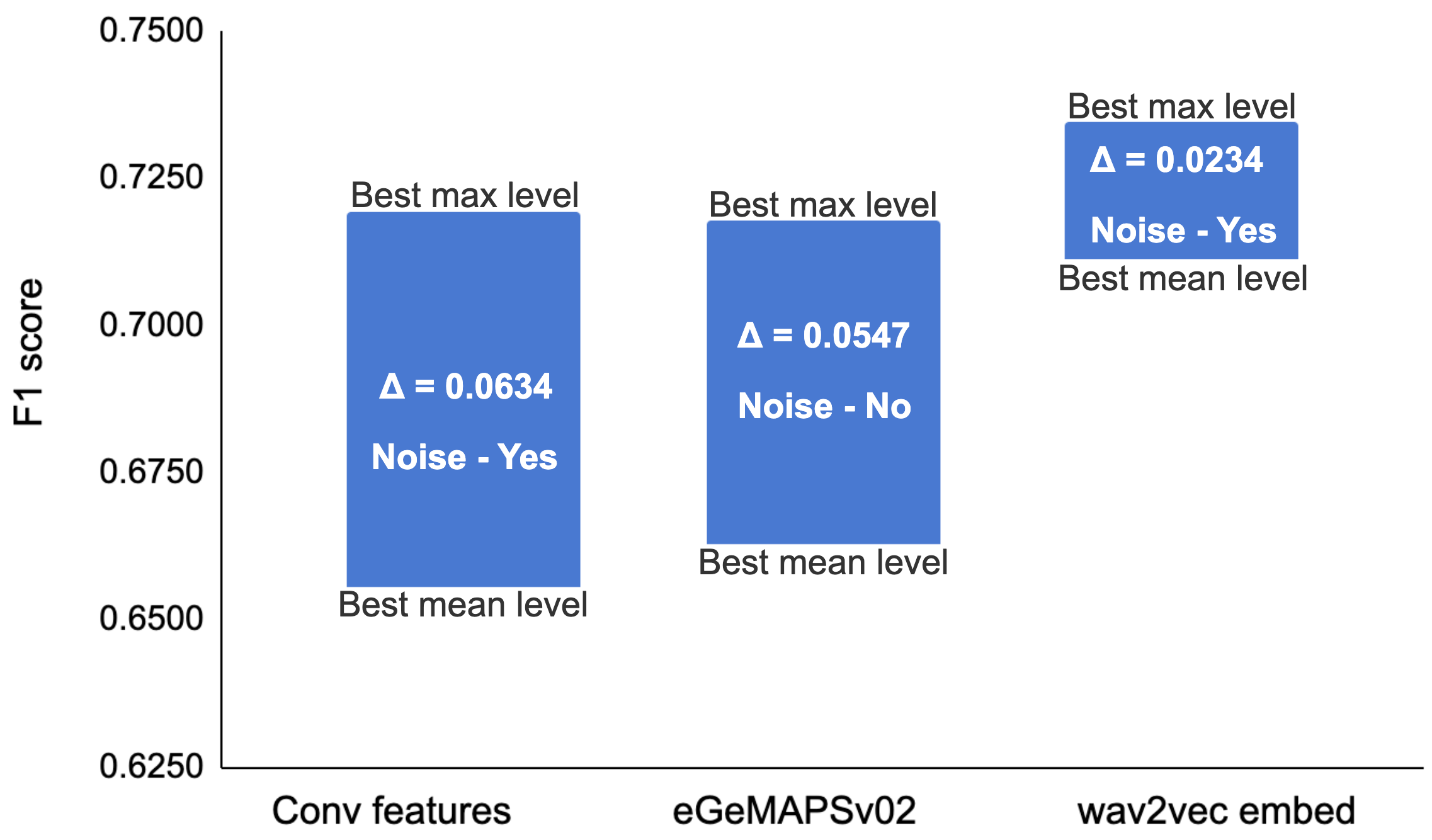}
    \caption{AD classification performance by feature type.}
    \label{fig:f1-features}
\end{figure}

The best mean F1 score represents the model that performs the best on average (across three random seeds) for some specific noise subcategory. Based on the best mean F1 score, the \wav-based model outperforms substantially the \gemaps-based model, while the \conv-based model achieves the lowest best mean level of performance (see Figure~\ref{fig:f1-features}). Interestingly in all three cases, the best mean level of performance is achieved by the models trained on the original audio without noise addition.

The best maximum F1 score represents the best possibly achievable performance across all the seeds, i.e. the model that performs the best on a single best seed. The difference between the best mean level and the best maximum level shows the potential of the models to achieve higher level of performance. Figure~\ref{fig:f1-features} shows that such a potential is the strongest for the \conv-based models (+6.3\%), and there is not that much room for improvement for the \wav-based models (+2.3\%). However, given the strong starting point, i.e. the strong best mean level, the absolute best maximum level of performance is achieved by the \wav-based model. Interestingly, this best maximum level is achieved by the model trained on the noisy data, not the original audio. The same is true for the second-best maximum performance, i.e. of the \conv-based model.

\subsubsection{Analysis Per Model Type}
\begin{figure}[t!]
    \centering
    \includegraphics[width = \linewidth]{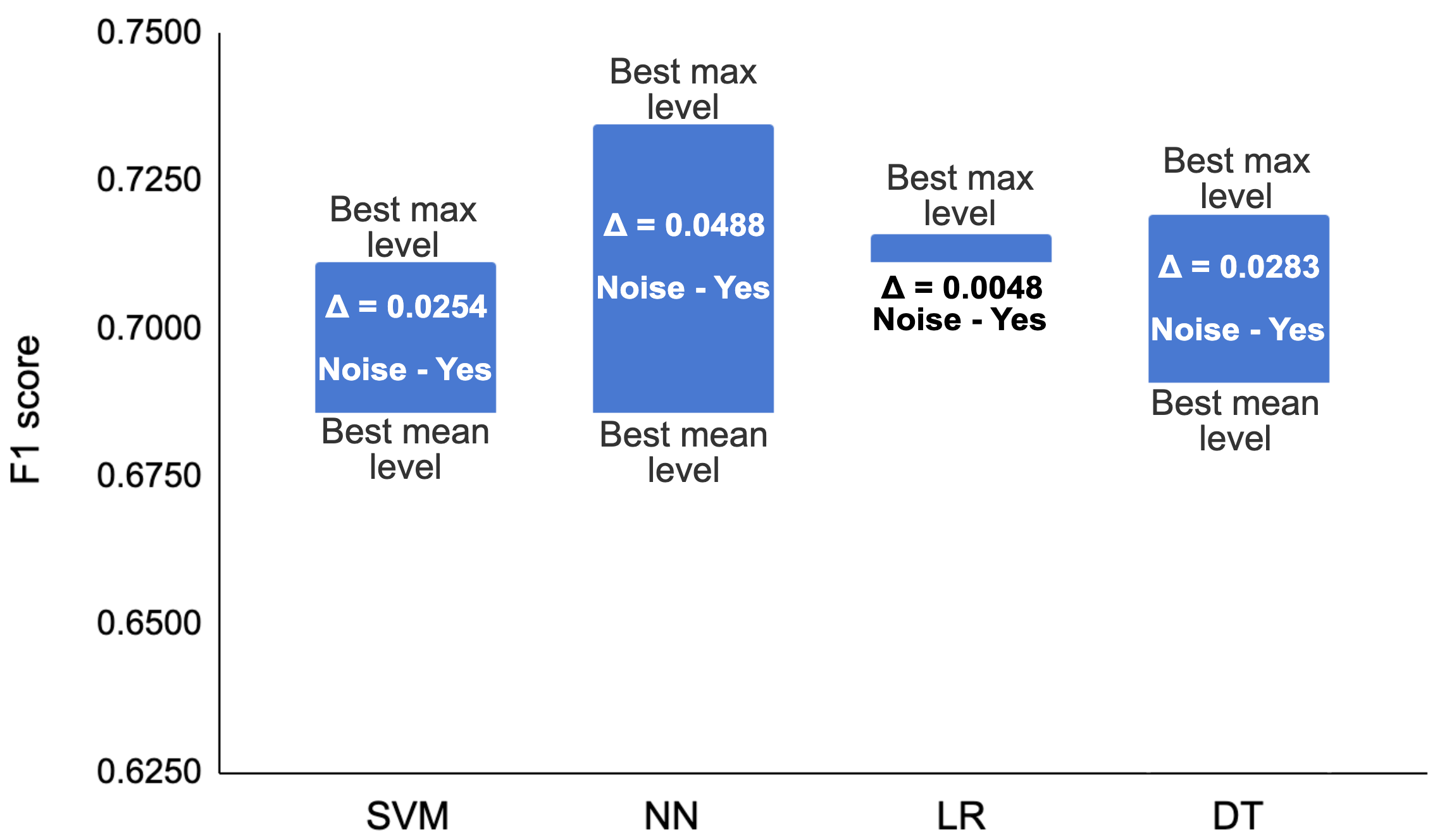}
    \caption{AD classification performance by model.}
    \label{fig:f1-models}
\end{figure}

The best mean F1 score is achieved by the LR model, while SVM and NN both share the lowest level of the best mean performance (Figure~\ref{fig:f1-models}). The growth potential of both linear models (LR and SVM) is weaker than that of the non-linear models (DT and NN), with the NN model showing the strongest potential across all model types. Once again, the best mean level of all the models is achieved when training the models on the original noise-free recordings, while the best maximum level is always achieved by training the models on the noisy audio recordings.

To overview, the results strongly suggest that noise has a beneficial effect on performance of AD classifiers, both linear and non-linear and utilizing different sets of features. However, all these performance results are aggregated across different categories and subcategories of noise. Next, we investigate in more detail how each specific noise category affects AD classification model performance.

\begin{table*}[t!]
\begin{adjustbox}{max width=1.\linewidth, center}
\begin{tabular}{llllcccc}
\textbf{\begin{tabular}[c]{@{}l@{}}Noise\\category\end{tabular}} & \textbf{Subcategory} & \textbf{Features} & \textbf{Count} & \textbf{\begin{tabular}[c]{@{}l@{}}Mean F1\\w/ noise\end{tabular}} & \textbf{\begin{tabular}[c]{@{}l@{}}Max F1\\w/ noise\end{tabular}} & \textbf{\begin{tabular}[c]{@{}l@{}}Best mean F1\\w/o noise\end{tabular}} & \textbf{\begin{tabular}[c]{@{}l@{}}Best max F1\\w/o noise\end{tabular}} \\
\hline \hline
\multirow{8}{*}{Animals} & \multirow{3}{*}{cat} & \conv & 24 & 0.6232 & \textbf{0.6842*} & 0.6557 & 0.6557 \\
 &  & \gemaps & 24 & 0.6284 & \textbf{0.6907*} & 0.6557 & 0.6557 \\
 &  & \wav & 24 & 0.6222 & 0.7006 & 0.7111 & \textbf{0.7200} \\
 \cdashline{2-8}
 & \multirow{3}{*}{crow} & \conv & 24 & 0.6217 & \textbf{0.6591} & 0.6557 & 0.6557 \\
 &  & \gemaps & 24 & 0.6345 & \textbf{0.6800*} & 0.6557 & 0.6557 \\
 &  & \wav & 24 & 0.6400 & 0.6878 & 0.7111 & \textbf{0.7200*} \\
  \cdashline{2-8}
 & \cellcolor{green!30}\multirow{2}{*} & \conv & 24 & 0.6255 & \textbf{0.6704} & 0.6557 & 0.6557 \\
 & \cellcolor{green!30}{dog} & \gemaps & 24 & 0.6318 & \textbf{0.7014*} & 0.6557 & 0.6557 \\
 \hline
\multirow{7}{*}{Natural} & \multirow{3}{*}{\begin{tabular}[c]{@{}l@{}}chirping\\birds\end{tabular}} & \conv & 24 & 0.6273 & \textbf{0.7018*} & 0.6557 & 0.6557 \\
 &  & \gemaps & 24 & 0.6443 & \textbf{0.6995*} & 0.6557 & 0.6557 \\
 &  & \wav & 24 & 0.6275 & 0.6966 & 0.7111 & \textbf{0.7200*} \\
  \cdashline{2-8}
 & \cellcolor{green!30}\multirow{2}{*} & \conv & 24 & 0.6229 & \textbf{0.6882*} & 0.6557 & 0.6557 \\
 & \cellcolor{green!30}{rain} & \gemaps & 24 & 0.6506 & \textbf{0.7135*} & 0.6557 & 0.6557 \\
  \cdashline{2-8}
 & \multirow{2}{*}{wind} & \conv & 24 & 0.6156 & 0.6480 & \textbf{0.6557} & \textbf{0.6557} \\
 &  & \gemaps & 24 & 0.6138 & \textbf{0.7019*} & 0.6557 & 0.6557 \\
 \hline
\multirow{7}{*}{Human} & \cellcolor{green!30}\multirow{2}{*} & \conv & 24 & 0.6182 & \textbf{0.6923*} & 0.6557 & 0.6557 \\
 & \cellcolor{green!30}{coughing} & \gemaps & 24 & 0.6387 & \textbf{0.7120*} & 0.6557 & 0.6557 \\
  \cdashline{2-8}
 & \cellcolor{green!30}\multirow{3}{*} & \conv & 24 & 0.6182 & \textbf{0.6816*} & 0.6557 & 0.6557 \\
 & \cellcolor{green!30}{crying baby} & \gemaps & 24 & 0.6472 & \textbf{0.7079*} & 0.6557 & 0.6557 \\
 & \cellcolor{green!30} & \wav & 24 & 0.6344 & \textbf{\textit{0.7345*}} & 0.7111 & 0.7200 \\
  \cdashline{2-8}
 & \multirow{2}{*}{sneezing} & \conv & 24 & 0.6071 & 0.6444 & \textbf{0.6557} & \textbf{0.6557} \\
 &  & \gemaps & 24 & 0.6406 & \textbf{0.6800*} & 0.6557 & 0.6557 \\
 \hline
\multirow{7}{*}{\begin{tabular}[c]{@{}l@{}}Domestic/\\interior\end{tabular}} & \multirow{2}{*}{\begin{tabular}[c]{@{}l@{}}clock\\ticking\end{tabular}} & \conv & 24 & 0.6013 & \textbf{0.6557} & \textbf{0.6557} & \textbf{0.6557} \\
 &  & \gemaps & 24 & 0.6284 & \textbf{0.6990*} & 0.6557 & 0.6557 \\
  \cdashline{2-8}
 & \multirow{2}{*}{\begin{tabular}[c]{@{}l@{}}vacuum\\cleaner\end{tabular}} & \conv & 24 & 0.5775 & 0.6292 & \textbf{0.6557} & \textbf{0.6557} \\
 &  & \gemaps & 24 & 0.5937 & \textbf{0.6561} & 0.6557 & 0.6557 \\
  \cdashline{2-8}
 & \multirow{3}{*}{\begin{tabular}[c]{@{}l@{}}washing\\machine\end{tabular}} & \conv & 24 & 0.6254 & \textbf{0.6919*} & 0.6557 & 0.6557 \\
 &  & \gemaps & 24 & 0.6391 & \textbf{0.6990*} & 0.6557 & 0.6557 \\
 &  & \wav & 24 & 0.6194 & 0.6816 & 0.7111 & \textbf{0.7200*} \\
 \hline
\multirow{8}{*}{\begin{tabular}[c]{@{}l@{}}Urban/\\exterior\end{tabular}} & \multirow{3}{*}{car horn} & \conv & 24 & 0.6324 & \textbf{0.7111*} & 0.6557 & 0.6557 \\
 &  & \gemaps & 24 & 0.5868 & \textbf{0.6832} & 0.6557 & 0.6557 \\
 &  & \wav & 24 & 0.6069 & 0.6631 & 0.7111 & \textbf{0.7200*} \\
  \cdashline{2-8}
 & \multirow{3}{*}{siren} & \conv & 24 & 0.6274 & \textbf{0.7191*} & 0.6557 & 0.6557 \\
 &  & \gemaps & 24 & 0.6098 & \textbf{0.6919*} & 0.6557 & 0.6557 \\
 &  & \wav & 24 & 0.5961 & 0.6667 & 0.7111 & \textbf{0.7200*} \\
  \cdashline{2-8}
 & \cellcolor{green!30}\multirow{2}{*} & \conv & 24 & 0.6112 & \textbf{0.6818*} & 0.6557 & 0.6557 \\
 & \cellcolor{green!30}{train} & \gemaps & 24 & 0.6382 & \textbf{0.6866*} & 0.6557 & 0.6557
\end{tabular}
\end{adjustbox}
\caption{Change in AD classification performance when models are trained on the noisy audio recordings, by noise category, subcategory and feature type. \textbf{Bold} denotes best performance per noise subcategory+features, \textbf{\textit{bold italic}} denotes best overall performance, \colorbox{green!30}{green background} denotes noise subcategory that has consistently highest performance when models are trained on the noisy recordings. * indicates significant difference of $p<0.05$ on McNemar's test.}
\label{tab:noise-category-perf}
\end{table*}

\subsubsection{Analysis Per Noise Type} 
The results of classification experiments with models trained on the noise-free and noisy audio show that best average classification performance is achieved when models are trained on clean noise-free audio recording (\textit{Best mean F1 w/o noise} and \textit{Mean F1 w/ noise} columns in Table~\ref{tab:noise-category-perf}). However, the maximum performance is consistently higher for the models trained on the noisy audio (columns \textit{Max F1 w/ noise} vs \textit{Best max F1 w/o noise} in Table~\ref{tab:noise-category-perf}). 

\begin{table*}[t!]
\begin{adjustbox}{max width=1.\linewidth, center}
\begin{tabular}{ll|ccc|ccc|ccc}
  &  & \multicolumn{3}{c|}{\textbf{Original noise-free audio}} & \multicolumn{3}{c|}{\textbf{Short noise}} & \multicolumn{3}{c}{\textbf{\textbf{Background noise}}} \\
 \textbf{F1} & \textbf{Model} & \conv & \gemaps & \wav & \conv & \gemaps & \wav & \conv & \gemaps & \wav \\
 \hline \hline
\multirow{4}{*}{Mean} & SVM & 0.6557 & 0.6480 & 0.6857 & \cellcolor{green!30}0.6816 & \cellcolor{green!30}0.6484 & \cellcolor{green!30}0.6885 & 0.6096 & \cellcolor{green!30}\textbf{0.7079} & 0.6067 \\
 & LR & 0.5698 & 0.6630 & 0.7111 & \cellcolor{green!30}0.6441 & 0.6413 & \cellcolor{green!30}\textbf{0.7159} & \cellcolor{green!30}0.6243 & 0.6369 & 0.5914 \\
 & NN & 0.6289 & 0.6541 & 0.6857 & \cellcolor{green!30}0.6355 & \cellcolor{green!30}0.6603 & \cellcolor{green!30}\textbf{0.7061} & 0.6206 & \cellcolor{green!30}0.6595 & 0.5901 \\
 & DT & 0.5882 & 0.5567 & \textbf{0.6908} & \cellcolor{green!30}0.6142 & \cellcolor{green!30}0.6004 & 0.6113 & 0.5154 & \cellcolor{green!30}0.6234 & 0.5653 \\
 \hline
\multirow{4}{*}{Max} & SVM & 0.6557 & 0.6480 & 0.6857 & \cellcolor{green!30}0.6816 & \cellcolor{green!30}0.6484 & \cellcolor{green!30}0.6885 & 0.6096 & \cellcolor{green!30}\textbf{0.7079} & 0.6067 \\
 & LR & 0.5698 & 0.6630 & 0.7111 & \cellcolor{green!30}0.6441 & 0.6413 & \cellcolor{green!30}\textbf{0.7159} & \cellcolor{green!30}0.6243 & 0.6369 & 0.5914 \\
 & NN & 0.6519 & 0.7177 & 0.7200 & \cellcolor{green!30}0.6705 & 0.6832 & \cellcolor{green!30}\textbf{0.7345} & \cellcolor{green!30}0.6484 & 0.6634 & 0.6034 \\
 & DT & 0.6250 & 0.5795 & \textbf{0.7045} & \cellcolor{green!30}0.6292 & \cellcolor{green!30}0.6077 & 0.6328 & 0.5263 & \cellcolor{green!30}0.6292 & 0.5843
\end{tabular}
\end{adjustbox}
\caption{Classification performance of the models trained on the noisy audio recordings with the sounds of crying baby. Mean F1 is averaged across three random seeds. \textbf{Bold} denotes the best performance per noise type (\textit{short} and \textit{background}), \colorbox{green!30}{green background} denotes performance that is higher than the analogous one for the model+feature set trained on the original noise-free audio.}
\label{tab:baby-crying}
\end{table*}

Out of all the noise categories, domestic/interior sounds seem to be the least beneficial for the AD classification models - none of the noise subcategories helps consistently improving classification performance. In the other categories, such as animal sounds, natural sounds, and urban/exterior noise, at least one noise subcategory consistently achieves substantially higher performance with the models trained on the noisy recordings, with all the tested audio features. The human noise is the most beneficial noise category for getting high AD classification results: 1) the overall best classification performance is achieved by the model trained on the noisy recording of this category (model trained on wav2vec embeddings of the audio with the \textit{crying baby} noise), 2) two out of three noise subcategories (\textit{coughing} and \textit{crying baby}) consistently achieve higher performance level across all the audio features. The best overall performance motivates us to investigate in more detail the classification performance of the models trained on the audio with the \textit{crying baby} noise.

\subsubsection{Analysis of the \textit{Crying Baby} Noise} 
All the \conv-based models  trained on the audio recordings with the sounds of \textit{crying baby} present as short noise, perform better than those same models trained on the original noise-free audio recordings (see Table \ref{tab:baby-crying} for details). Same is true for the majority of \wav-based models, with \wav-based NN achieving the overall best performance. 

When it comes to the sound of crying baby to be introduced as a continuous background noise, the overall performance level of \wav and \conv-based models decreases substantially. \wav-based models are not able anymore to outperform any of noise-free models, and only linear \conv-based models are still able to outperform their noise-free analogues. The \gemaps-based SVM model is able to achieve its best performance with this type of noise. 

\subsection{Recommendations}

Based on the results of our analysis, we outline a set of recommendations for the ML researchers and practitioners interested in deploying AD classification models in real world. 

First, if acoustic features are extracted using conventional and not deep learning-based features, such as \conv or \gemaps, it is important to use the noise removal speech pre-processing techniques to normalize the audio dataset that is used for training ML models. As explained in Section~\ref{sec:noise-effect-values}, even short segments of unwanted noise, such as accidental siren, craw caw or a short vacuum cleaner sound, may significantly change more than 50\% of acoustic features. Having the training dataset where otherwise similar datapoints are represented by significantly different acoustic features, introduces many unnecessary challenges in model development.

Second, it is important to make sure the deployed models are not be used in certain types of real world environments where certain noises are common. As explained in Section~\ref{sec:noise-clf-perf}, domestic noise, such as washing machine or vacuum cleaner, may decrease classification performance. As such, it is important to recommend the real world users of the AD classification model to avoid this type of noise when recording audio in order to expect better accuracy of the model. Other noises, such as baby cry, cough or dog bark, are not harmful and there is no need to avoid them. This is also important to know because these types of noise are much more difficult to securely avoid in real world scenarios than sounds of a vacuum cleaner or washer. 

Third, model developers should expect different effects of noise on the AD classification performance depending on the type of audio representation and model used. Deep features, such as \wav, are affected less strongly by the presence of noise comparing to more conventional acoustic features, such as \conv and \gemaps, although models utilizing all three types of features may benefit from certain noises in audio. More simplistic linear models, such as SVM and LR, may be impacted positively but not very strongly (up to 2.5\%) by the presence of appropriate noise in the recordings. The more complex non-linear models, such as DT and NN, may experience twice stronger positive effect (+4.8\%) due to appropriate noise.

\section{Conclusions}

In this paper, we study the effect of fifteen types of acoustic noise, standard in home environments, on AD classification from speech. We perform a thorough analysis showing how four ML models and three types of acoustic features are affected by different types of acoustic noise. We show that natural environmental noise is not necessarily harmful, with certain types of noise even being beneficial for AD classification performance and helping increase accuracy by up to 4.8\%. We provide recommendations on how to utilize
acoustic noise in order to achieve the best performance results with the ML models deployed in real world in order to facilitate the use of scalable digital health tools for AD detection from speech. Further research is necessary to investigate the effect of more types of acoustic noise common in real world scenarios.

\newpage
\bibliography{anthology,custom}
\bibliographystyle{acl_natbib}

\end{document}